\documentstyle[12pt]{article}
\begin{document}
\titlepage

\title{\begin{flushright}
Preprint PNPI-2454,  2001
\end{flushright}
\bigskip\bigskip\bigskip
Using of unitarity equations for the
calculation of  fermion interaction amplitudes
in the superstring theory}
\author{G.S. Danilov \thanks{E-mail address:
danilov@thd.pnpi.spb.ru}\\ Petersburg Nuclear Physics
Institute,\\ Gatchina, 188300, St.-Petersburg, Russia}

\maketitle
\begin{abstract}
The unitarity equations for the boson interaction amplitudes in
the superstring theory  are used to calculate the
interaction amplitudes including the Ramond states, which are
10-fermion and Ramond bosons. The n-loop, 4-point amplitude with
two massless Neveu-Schwarz bosons and two massless Ramond
states is obtained explicitly.
It is shown that, in addition, the
unitarity equations require some integral relations for local
functions determining the amplitude.
For the tree  amplitude the validness of the above
integral relations is verified.

\end{abstract}
\newpage

\section {Introduction}

A calculation of interaction amplitudes with Ramond
state legs (they are 10-dimensional fermions and Ramond bosons)
for the Ramond-Neveu-Schwarz superstring \cite{gsw} looks
\cite{gsw,fried,nil} as a very complicated task.
We propose to calculate the desired amplitudes
from the unitarity equations for amplitudes of the interaction
of the Neveu-Schwarz boson states.  In this case the desired
$n$-loop amplitude is given by integral of an expression, which
is represented through measures and vacuum correlators on the
genus-$n_1>n$ supermanifold where certain handles are
degenerated.  Now the measures and vacuum correlators are known
\cite{danphr,prepr,dannph,dan96} for any genus and for all spin
structures, the Ramond sector being included.

To obtain the desired representation of the Ramond state
interaction amplitude, we calculate the fermion loop
discontinuity in the center mass energy (is more exact, in a
relevant by-linear 10-invariant) of the boson state amplitude.
The discontinuity is
represented by an integral of by-linear product of
the amplitudes with fermions in the intermediate state.
As far as
Ramond states are constructed using Majorano-Weyl spinors
\cite{gsw}, the amplitude contains only odd number of Dirac
matrices sandwiched by spinors.  A number of the above spinor
structures is less, than number of
conditions, which is possible to satisfy locally when
the discontinuity
is represented
in the form of the {\it trace} of Dirac
matrix products. Thus, the unitarity equations  require, in
addition, certain integral relations to be valid.

The above calculation use some
general properties of integration measures and vacuum
correlators when certain handles are degenerated. Details
of expressions given in \cite{danphr,prepr,dannph,dan96}  are not
cruel for the calculation.

In the present paper we calculate the $n$-loop, 4-point
amplitude of two massless NS bosons and two massless Ramond
states. The $n$-loop amplitude is
represented by the integral of local functions determining the
integrand for the $(n+1)$-loop massless boson scattering
amplitude
in the region where one of the handles is degenerated.

The paper is organized as follows.
In section 2 expression for the boson emission amplitude is
given along with some formulas necessary for the following
consideration. In section 3 the two-particle unitarity equations
are derived. In section 4 the unitarity equations due the boson
loop are considered. In section 5 the fermion loop unitarity
equations are discussed. The 4-point amplitude with two NS
massless bosons ant two massless Ramond states is calculated for
any number of loops. Additional integration relations are
obtained, which are necessary to  provide the unitarity
equations.  The tree approximated relations
are verified.

\section {Expressions for amplitudes}
The superstring
amplitude with number of loops $n$ is given by integral of
the sum over super-spin structures. The above super-spin
structures are defined on genus-$n$ complex $(1|1)$
supermanifold \cite{bshw} by the super-Schottky groups
\cite{danphr,dannph,dan3}.  For $2\pi$-twist
around $B_s$-cycle the corresponding transformation  is
determined by the Schottky multiplier $k_s$ along with  two
fixed points $U_s=(u_s|\mu_s)$ and $V_s=(v_s|\nu_s)$ on the
supermanifold.  Here $\mu_s$ and $\nu_s$ are Grassmann partners
of local points $u_s$ and,respectively, $v_s$. For the boson
handle (the NS case), to $2\pi$-twist around $A_s$-cycle the
identical transformation corresponds. For the Ramond type handle
$2\pi$-twist around $A_r$-cycle is performed by the
transformation with $k_r=1$ and $\sqrt k_r=-1$, the fixed points
being the same as for the transformation assigned to
$2\pi$-twist about $B_r$-cycle \cite{danphr,dannph,dan3}.  Thus,
to each handle, $(3|2)$ complex parameter correspond to be
$(k_s,u_s,v_s)$ and $(\mu_s,\nu_s)$. The superstring amplitude
with number of loops $n $ is given by integral
\cite{danphr,dannph} over $(k_s,u_s,v_s, \mu_s,\nu_s)$ and their
complex conjugated, and over interaction vertex coordinates
$t_j=(z_j|\vartheta_j)$  on the supermanifold. In doing so the
$\{N_0\}$ set of any $(3|2)$ complex variables  is fixed due to
$SL(2)$ symmetry that leads to an additional factor
$|H(\{N_0\})|^2$ in the integrand. The above factor is given in
\cite{danphr}.  For each superspin structure in
amplitude of interaction $m$ bosons, the integrand is the
product \footnote{The overline denotes the complex conjugation.}
of the integration measure $Z_{L,L'}^{(n)}(\{q,\overline q\})$
by the vacuum expectation $F_m^{(n)}(\{t_j,\overline
t_j\},\{p_j, \epsilon^{(j)}\},\{q,\overline q\};L,L')$ of the
interaction vertex product. Here $t_j=(z_j|\vartheta_j)$ is the
coordinate of the $j$-th boson vertex and $\{q\}$ is the set the
super-Schottky group parameters. Further, $p_j$ and
$\epsilon^{(j)}$  is the momentum and, respectively, the
polarization tensor of the $j$-th boson.  Furthermore, $L$
($L'$) is the superspin structure for the holomorphic
(anti-holomorphic) movers.  Like usual spin structure \cite{sw},
the super-spin one is given by theta-function characteristics
$l_{1s}$ and $l_{2s}$ assigned to the given handle $s$. In doing
so $l_{1s}$ and $l_{2s}$ can be restricted by zero or
$1/2$. For the boson loop $l_{1s}=0$. For the fermion one
$l_{1s}=1/2$ ($l_{2s}=0$ and $l_{1s}=1/2$
corresponds to the loop of even spin structure and,
respectively, of the odd spin one). The $n$-loop amplitude
$A_m^{(n)} (\{p_j,\epsilon^{(j)}\})$ for interaction of $m$
bosons is given by
\begin{eqnarray}
A_m^{(n)}(\{p_j\},\{\epsilon^{(j)}\})=\frac{g^{2n+2}}{2^nn!}\int
|H(\{N_0\})|^2\sum_{L,L'}
Z_{L,L'}^{(n)}(\{q,\bar q\})
\nonumber\\
\times
F_m^{(n)}
(\{t_j,\overline t_j\},\{p_j\},\{\epsilon^{(j)}\},\{q,\bar
q\};L;L')(dqd\bar qdtd\bar t)'
\label{ampl}
\end{eqnarray}
where $g$ is a coupling constant, and other
definitions have been explained above.  The integration is
performed over $(q,\bar q)$  and over the interaction vertex
coordinates $(t,\bar t)$ with
$t=(z|\vartheta)$. Furthermore, $(3|2)$ complex variables are
fixed due to $SL(2)$-symmetry.  For the 4-point massless boson
amplitude only even super-spin structures
present (including those which contain even number of the
fermion loops of the odd spin structure). The multiplier $1/2^n$
is due to the symmetry under the $U_s\rightleftharpoons V_s$
interchange of fixed points $U_s$ and $V_s$ for the given
handle, and $1/n!$ appears due to the symmetry under the
interchanging of the handles.  For any boson variable $x$ we
define $dxd\overline x=d(Re\,x)d(Im\,x)/(4\pi)$.  For any
Grassmann variable $\eta$ we  define $\int d\eta\eta=1$.  The
tension of a string is equal $1/\pi$. The normalizations
are as in \cite{gsw}.

Due to factorization in left, right movers \cite{bk}, the
integration measure is given by
\cite{danphr}:
\begin{equation}
Z_{L,L'}^{(n)}(\{q,\overline
q\})=(8\pi)^{5n}[\det\Omega_{L,L'}^{(n)}(\{q,\overline q
\})]^{-5} Z_L^{(n)}(\{q\}) \overline {Z_{L'}^{(n)}(\{q\})}
\label{hol}
\end{equation}
where
$Z_L^{(n)}(\{q\})$  is a holomorphic function of
$\{q\}$, and $\Omega_{L,L'}^{(n)}
(\{q,\overline q\})$ is expressed through the period matrix
$\omega^{(n)}(\{q\},L)$:
\begin{equation}
\Omega_{L,L'}^{(n)}(\{q,\overline q\})=
2\pi
i[\overline{\omega^{(n)}(\{q\},L')}-
\omega^{(n)}(\{q\},L)].
\label{grom}
\end{equation}
Because of
the boson-fermion mixing the period matrix
depends \cite{pst,danphr} on $L $. The holomorphic
function in (\ref{hol}) is given by
\begin{equation}
Z_L^{(n)}(\{q\})=\tilde Z^{(n)}(\{q\},L)\prod_{s=1}^n
\frac{ Z^{(1)}(k_s;l_{1s},l_{2s})}{k_s^{(3-2l_{1s})/2}
(u_s-v_s-\mu_s\nu_s)}
\label{zinv}
\end{equation}
where the theta-characteristics $(l_{1s},l_{2s})$
are equal 0 or 1/2, function $\tilde Z^{(n)}(\{q\},L) $
is given in \cite{danphr}, and
\begin{equation}
Z^{(1)}(k;l_{1},l_{2})=(-1)^{2l_{1s}+2l_{2s}}16^{2l_{1s}}
\prod_{p=1}^\infty
\frac{[1+(-1)^{2l_2}k^pk^{(2l_1-1)/2}]^8}{[1-k^p]^8}\,.
\label{z1h}
\end{equation}
As it was already mentioned, $F_m^{(n)}(\{t_j\},\{p_j\},
\{\epsilon^{(j)}\},\{q,\bar q\};L;L')$ in (\ref{ampl})
is the vacuum expectation of the interaction vertex product. The
above vertex for the massless boson
with a 10-momentum $p =\{p_M\}$ and
polarizations $\zeta=\{\zeta_M\}$ and $\zeta'=
\{\zeta_M'\}$ for left and right movers
($M=0,\dots,9 $), is given
(with our normalization) by \cite{fried}
\begin{equation}
V(t,\overline
t;p;\zeta)=4[\zeta D(t)X(t,\overline
t)][\overline{\zeta'D(t)}X(t,\overline t)]
\exp[ipX(t,\overline
t)]
\label{vert}
\end{equation}
where for two
10-vectors $a$ and $b$ the scalar product $ab=a_Mb^M $
is defined using the " mainly $+$ "
metrics.  Through $X^N (t,\bar t)$ the scalar superfield of the
string is denoted.  Further, $p\zeta=p\zeta'=0$, $p^2=0$,
and the spinor derivative $D (t) $ is
\begin{equation}
D(t)=\vartheta\partial_z+\partial_\vartheta
\label{supdr}
\end{equation}
Function $F_m^{(n)}(\{t_j,\overline t_j\},\{p_j\},
\{\epsilon^{(j)}\},\{q,\bar q\};L;L')
\equiv F_m^{(n)}(\{t_j,\overline t_j\},\{p_j\},
\{\epsilon^{(j)}\})$ in (\ref{ampl})  is
given \cite{gsw} through the scalar superfield vacuum correlator
$\hat X_{L,L'}(t_j,\overline t_j;t_l,\overline t_l;\{q\})
\equiv\hat X(j,l)$ by the integral over
Grassmann variables $\{\eta_j,\overline\eta_j\}$ assigned to the
$j$-th boson
\begin{equation}
F_m^{(n)}(\{t_j,\overline
t_j\},\{p_j\},\{\epsilon^{(j)}\}) =\int(d\eta d\bar\eta)
\exp\Biggl[-\frac{1}{2}\sum_{j,l} (\hat\kappa_j+ip_j)
(\hat\kappa_l+ip_l) \hat X(j,l)\Biggl] \label{funct}
\end{equation}
where the operator $
\hat\kappa_j$ is  $\hat\kappa_j=\kappa_j
+\overline{\kappa_j'}$ with $
\kappa_j=2\eta_j\zeta_{(j)}D(t_j)$ and $
\kappa_j'=2\eta_j\zeta_{(j)}'D(t_j)$.  Here $\zeta_{(j)}$
and $\overline{\zeta_{(j)}'}$ determine the
polarization of the $j $-boson. So $\zeta_{(j)M}\overline
{\zeta_{(j)N}'}= \epsilon_{MN}^{(j)}$.  All particle momenta are
implied to be "entering" to diagram.

The vacuum correlator is given through  Green
function $R_L^{(n)}(t,t';\{q\})$ along with the super-scalar
functions $J_s^{(n)}(t;\{q\};L)$, having periods ($s=1,
\dots, n $):
\begin{equation}
4\hat X_{L,L'}(t,\overline t;t',\overline t';\{q\})=
R_L^{(n)}(t,t';\{q\})+\overline{R_{L'}^{(n)}(t,t';\{q\})}
+I_{LL'}^{(n)}(t,\overline
t;t',\overline t';\{q\})\,,
\label{corr}
\end{equation}
\begin{eqnarray}
I_{LL'}^{(n)}(t,\overline
t;t',\overline t';\{q,\bar q\})= [J_s^{(n)}(t;\{q\};L) +
\overline{J_s^{(n)}(t;\{q\};L')}]
[\Omega_{L,L'}^{(n)}(\{q,\overline q \})]_{sr}^{-1}
\nonumber\\ \times
[J_r^{(n)}(t';\{q\};L)+\overline{J_r^{(n)}(t';\{q\};L')}]
\label{illin}
\end{eqnarray}
Green's function is normalized by
\begin{equation}
R_L^{(n)}(t,t';\{q\})=\ln(z-z'-\vartheta\vartheta')+
\tilde R_L^{(n)}(t,t';\{q\})
\label{lim}
\end{equation}
where $\tilde R_L^{(n)}(t,t';\{q\})$ has no a singularity at
$z=z'$.  At $t=t'$, as is usual,
$R_L^{(n)}(t,t';\{q \})$ is $
\tilde R_L^{(n)}(t,t;\{q\})$. The correlator (\ref{corr}) at the
same points is calculated accordingly to this agreement.

\section {Two-particle  unitarity equations}

The domain where certain Schottky
multipliers $|k_j|\to1$ can be excluded \cite{prepr}
from the
integration region in (\ref{ampl})
Then the unitarity equations are due
\cite{gsw} to the region where
certain $k_j\to0$.
Two-particle
cut is due to the region where only Schottky multiplier
$k\to0 $. In that case in (\ref{ampl}) it is convenient to fix
local coordinate of one of the interaction vertices, for example,
$z_4$, along with limiting points $U= (u|\mu)$ and
$V=(v|\nu)$ of the degenerated transformation discussed.
Moreover, we take $\mu=\nu=0$.
Then $|H(\{N_0\})|^2$ in (\ref{ampl}) is given by
\cite{danphr}:
\begin{equation}
|H(\{N_0\})|^2=|(z_4-u)(z_4-v)|^2\,.
\label{fix}
\end{equation}
We calculate the discontinuities in $s
=-(p_1+p_2)^2=-(p_3+p_4)^2$.  They arise from the region
where both $z_1$ and $z_2$ being outside the Schottky circles,
go to $u$ or to $v$.
The  contribution to integral  from the $z_2\to v$ region is
the same as from the $z_2\to u$ one. So, we consider  $z_2\to
u$, the result being multiplied by factor 2. Besides, the result
needs to be multiplied by factor $n$ since each of $n$ handles
can be the degenerated handle discussed.

If $z_2\to u$, then $z_1\to u$ or $z_1\to v$. The second case
is reduced to the first one by the appropriate
Schottky transformation of $z_1 $.  Thus $z_1 $
appears inside the circle containing $u$. Therefore we
consider $z_1\to u$ and $z_2\to u$ for $z_1 $ can be both
outside the Schottky circle, and inside it, and $z_2 $ lays
outside of the Schottky one.  In that case
\begin{equation}
|z_2-u|\geq\sqrt k|u-v|\,.
\label{bound}
\end{equation}
This formula follows from  an expression \cite{danphr,dannph}
of the Schottky transformation through
$u$, $v$ and $k$ at $k\to0$. For simplicity we
calculate that part of the discontinuity, which contains
the tree amplitude by the $(n-1)$-loop one. It
is determined by a configuration where limiting points
of remaining basic Schottky group transformations are
not closed to $u$. We restrict ourselves by the discussion
of the massless states.

The singularity at $k\to0 $ of the integration measure
(\ref{hol}) arises due to factor $1/k^{(3-2l_1)/2} $ in
(\ref{grom}). The leading singularity $\sim1/k^{3/2} $ for
the NS handle (then $l_1=0 $, see section 2)
is cancelled for the sum over two terms  $l_2=0 $ and
$l_2=1/2 $ since (\ref{ampl}) is even function
of $\sqrt k$. So the leading singularity in integral
always is $\sim1/|k|^2$. The factor at $1/|k|^2$
depends $\ln|k|$ by means of the period
matrix (see the Appendix).  Next order in degrees of $k$ terms do
not contribute to the massless state cuts considered.
With the required accuracy we can present the contribution $W $
to (\ref{ampl}) from discussed region  as the following
integral over the loop momentum $\tilde p_1$  of the
sum of expressions factorized in initial and final states
\begin{equation}
W=\int\frac{2d^{10}\tilde p_1}{\tilde p_1^2\tilde p_2^2}
\sum_{(\lambda,\lambda')}
\hat A_{(0)}(\{p,\epsilon\}_{(i)},\tilde p_1
\lambda,\lambda')\hat A_{(n-1)}(\lambda,\lambda';\tilde p_1,
\{p,\epsilon\}_{(f)})\,,
\label{unit}
\end{equation}
where $\tilde p_1+\tilde p_2=P=p_1+p_2 =-(p_3+p_4)$ and the
sum goes over polarizations $(\lambda,\lambda')$
of the intermediate state for holomorphic ($\lambda $) and
anti-holomorphic ($\lambda'$) movers.  The first
multiplier under the sum depends on momenta and
polarizations $\{p,\epsilon\}_{(i)}=
(p_1,p_2,\epsilon^{(1)},\epsilon^{(2)})$
of the initial state and of the intermediate state, as well.
The second one depends on parameters of
the intermediate state and on parameters
$\{p,\epsilon\}_{(f)}=(p_3,p_4,\epsilon^{(3)},\epsilon^{(4)})$
of the final state. Both multipliers are calculated for $\tilde
p_1^2=\tilde p_2^2=0$.  The cut due to the nullification of a
denominator $ (\tilde p_1^2\tilde p_2^2) $ is calculated without
an effort. From the comparison of eq.(\ref{unit}) with the
unitarity equation it follows that $\hat
A_{(0)}(\{p,\epsilon\}_{(i)}, \tilde p_1 \lambda,\lambda')$ and
$ \hat A_{(n-1)}(\lambda,\lambda';\tilde p_1,\{p,\epsilon
\}_{(f)}) $ are the tree and $(n-1)$-loop amplitudes.
To derive
(\ref{unit}) the considered contribution $W $ to (\ref{ampl})
is first represented by
\begin{eqnarray}
W= \int
d^{10}\tilde p_1\int \frac{d^2z_1}{4\pi}
\frac{d^2z_2}{4\pi}dy\frac{1}{2}
\exp\biggl[\frac{p_1p_2}{2}\ln|z_1-z_2|- \frac{p_1\tilde
p_1}{2}\ln|z_1-u|
\nonumber\\
-\frac{p_2\tilde p_1}{2}\ln|z_2-u|
+\frac{y\tilde p_1^2}{4}\biggl]
\sum_{r} \tilde
O_r(\{p,\epsilon,z,\bar z\}_{(i)},u,\bar u,\tilde p_1)
O_r(\{p,\epsilon\}_{(f)},\tilde p_1)\,,
\label{int}
\end{eqnarray}
where all of 10-vectors are Euclidian, and $y=\ln|k|$.
Function $\tilde O_r(\{p,\epsilon,t,\bar t\}_{(i)},u,
\bar u,\tilde p_1)$ has the translation symmetry. In addition, by
dimensional reasons, it receives multiplier $1/|\ell|^2$ if
any of $z_1 $ and $z_2 $ is multiplied by $\ell $.
Therefore, if, instead of $z_1 $, one defines a new variable $z
= (z_1-u)/(z_2-u)$, then the dependence on $z_2 $ and on $y $ is
extracted in (\ref{int}) in the form of the factor
$|z_2-u |^{-2}\exp[x(\tilde p_1^2-\tilde p_2^2)/4+
y\tilde p_1^2/4]$ where $x=\ln|z_2-u|$.
In this case $-\infty<y<0$ and, due to
(\ref{bound}), $0<x<y/2$. The integral over
$z_2 $ and $y $ being calculated, is $4/[\tilde
p_1^2(\tilde p_1^2+\tilde p_2^2)]$.  Since this expression
is further multiplied by the function,
which is symmetrical under $\tilde p_1\rightleftharpoons\tilde
p_2$, it can be replaced by $2(\tilde
p_1^2\tilde p_2^2)^{-1}$.  Then the required (\ref{unit})
appears. The tree amplitude arises in the form of
integral where interaction vertex coordinate for
intermediate particles with the $\tilde p_1$ momentum
is fixed to be $(0|0)$, and for the remaining intermediate state
it is
$(\infty|0)$.  In integral for the $(n-1)$-loop amplitude,
the vertex
coordinates corresponding to the above states are fixed to be
$(v|0)$ and $(u|0)$.  Local coordinate $z_4$ is fixed, too.  To
derive (\ref{int}), the vacuum expectation (\ref{ampl}) is
represented by the integral (\ref{funct}) over $ (d\eta
d\bar\eta) $.  Further, at $k\to0 $, every term $\hat W$ of the
integrand  is represented by the Gauss integral over the loop
momentum as it follows  \\ \\
\begin{eqnarray}
\hat W= \int
d^{10}\tilde p_1 \exp\Biggl[\tilde G_0+ \tilde G_1\tilde p_1^2
+\tilde B_1\tilde p_1\Biggl]
\exp\Biggl[G_1\tilde p_1^2
+G_{12}
\tilde p_1\tilde p_2
+G_2\tilde p_2^2+B_1\tilde p_1
\nonumber\\
+B_2\tilde p_2
+G_0
\Biggl]
O_{\tilde L}(\{p,\zeta\},\tilde p_1,\{t\},
\{q\})\overline{
O_{\tilde L'}(\{p,\zeta'\},\tilde p_1,\{t\},
\{q\})}
\nonumber\\
\times
\hat Z_{\tilde L,\tilde L'}^{(n-1)}(\{q,\bar q\}
|(z_4-u)(z_4-v)|^2\,,
\label{gint}
\end{eqnarray}
where
the integrand depends on the type of
the degenerated loop. Factors $\tilde G_0$,
$\tilde G_1$ and $\tilde B_1$ both depend on the initial
state, on $y=\ln|k|$ and on $(u,\bar u)$.
The second exponent depend on the final
state and on $\{q\}$ (with $k=0$ for the degenerated
handle). It depends also on super-spin structures $
(\tilde L,\tilde L')$ associated with  all handles,
except for the degenerated handle. In (\ref{gint}) the
multiplier (\ref{fix}) is given explicitly. Further,
$\{p\}$, $\{\zeta\}$, $\{\zeta'\}$ and $\{t,\bar t\}$
denote the sets of momenta, polarizations and vertex coordinates
for the bosons in the initial and final states.  Factors
$O_{\tilde L}(\{p\},\{\zeta\},\tilde p_1,\{t\},\{q\})$
and $\overline{O_{\tilde L'}(\{p\},\{\zeta'\},
\tilde p_1,\{t\},\{q\})}$ arise due to
an expansion in  small $k$, $(z_1-u)$ and
$(z_2-u)$ of the holomorphic or, respectively,
anti-holomorphic function in (\ref{hol}) and in (\ref{funct}).
Because of the non-holomorphic factor in (\ref{hol}) and due to
the last term in (\ref{corr}),
$O_{\tilde L}(\{p\},\{\zeta\},\tilde p_1,\{t\},\{q\})$
for $n > 1 $ depends also on $\{\bar t\}$, $\{\bar q\}$ and $L'$.
Correspondingly, the second of the discussed factors depends on
$\{t\}$, $\{\bar q\}$ and $L$.  Both the discussed multipliers
may be polynomial in $\tilde p_1$. Each of multipliers is the
sum of expressions factorized in the initial and final states.
Factor $\hat Z_{\tilde L,\tilde L'}^{(n-1)}(\{q,\bar q\}$
appears due to the integration measure. Below we obtain
expressions for the functions discussed. In deriving
(\ref{gint}) we shall use following formulas for a determinant
$n$-dimensional matrix $ \Omega$ and for the quadratic form
$J\Omega^{-1} J $ of $n$ functions $J_s $ (where $s=1, \dots,
n $):
\begin{eqnarray}
J\Omega^{-1}J=[J_1(1-\Omega_{1j_1}\Omega_{j_1l_1}^{-1})
J_{l_1}]^2
[\Omega_{11}-\Omega_{1j}\Omega_{jl}^{-1}\Omega_{l1}]^{-1}
+J_j[\Omega^{-1}]_{jl}J_l\,,
\nonumber\\
\det\Omega=\det\{\Omega_{jl}\}
[\Omega_{11}-\Omega_{1j_1}\Omega_{j_1l_1}^{-1}\Omega_{l_11}]\,,
\label{mast}
\end{eqnarray}
where the summation on
twice repeating indexes is implied, and alphabetic indexes in
(\ref{mast}) is run from 2 up to $n$. The coefficient at
$J_pJ_s$ (where $p,s=1,\dots,n$) in (\ref{mast}) are non other
than the appropriate element $(\Omega^{-1})_{ps}$ given through
elements of $\Omega$.  The second formula follows from equation
$ \ln\det\Omega=trace\ln\Omega $ once $trace\ln\Omega $ is
calculated in terms of quantities in (\ref{mast}).

\section {Unitarity for the boson loop}
In a case of the boson loop at
$k\to0$ the scalar function $J_1^{(n)}(t;\{q\};L)\equiv
J_1(t)$, as well as elements $\omega^{(n)}(\{q\},L)_{11}
\equiv\omega_{11}$ and $\omega^{(n)}(\{q\},L)_{1l}
\equiv\omega_{1l}$ matrices of the periods, are expressed
through holomorphic Green's function $R_{\tilde L}^{(n-1)}
(t,t';\{q\})\equiv R(t,t')$ and scalar functions
$J_1^{(n-1)}(t;\{q\};\tilde L)\equiv J_l(t)$ on the
genus-$(n-1)$ supermanifold
formed by other handles as follows:
\begin{eqnarray}
2\pi i\omega_{11}=\ln k+
R(U,U)+R(V,V)-R(U,V)-R(V,U)\,,
\nonumber\\
2\pi i\omega_{1l}=J_l(U)-J_l(V)\,,\quad
J_1(t)=R(t,U)-R(t,V)\,,
\label{rels}
\end{eqnarray}
where $U=(u|0)$ and $V=(v|0)$.
For the boson string and in
the NS sector of the superstring
\cite{div,vec,pst,dan89,dan0}
these relations directly follow
from a representation of the discussed functions by
Poincare series.
In particular, for the boson string the mentioned series
are given in \cite{danphr} (in its Appendix B).  For the NS
sector the discussed expressions are obtained by the replacement
of an interval $ (z_1-z_2) $ by a superinterval
$(z_1-z_2-\vartheta_1\vartheta_2)$. For zero Grassmann moduli
the relations are evidently true  for the Ramond sector,too.
Eqs.(\ref{rels}) are necessary for the unitarity equations
could be to take place. So it
is naturally expect that the relations is true even though the
Grassmann moduli are arbitrary, and it is really so in the case
(see the Appendix).

For simplicity we
assume the boson loop for both movers:
$l_1=l_1'=0$. Then exponents in
(\ref{gint}) are found to be:
\begin{eqnarray}
B_1=-i\sum_{j=3}^4(\hat\kappa_j+ip_j)
\hat X(t_j,\bar t_j;V,\bar V)\,,\,
B_2=-i\sum_{j=3}^4(\hat\kappa_j+ip_j)
\hat X(t_j,\bar t_j;U,\bar U)\,,
\nonumber\\
G_0=-\frac{1}{2}\sum_{j,l=3}^4(\hat\kappa_j+ip_j)
(\hat\kappa_l+ip_l) \hat X(t_3,\bar t_3;t_4,\bar t_4)\,,
\quad
G_{12}=\hat X(U,\bar U;V,\bar V)\,,
\nonumber\\
G_1=\hat X(V,\bar V;V,\bar V)+\ln|u-v|\,,
\quad
G_2=\hat X(U,\bar U;U,\bar U)\,,
\nonumber\\
\tilde G_0=-(\hat\kappa_1+ip_1)
(\hat\kappa_2+ip_2)\frac{1}{2}\ln|z_1-z_2-
\vartheta_1\vartheta_2|\,,\quad
\tilde G_1=\frac{1}{2}\ln|k|\,,
\nonumber\\
\tilde B_1=i\sum_{j=1}^2(\hat\kappa_j+ip_j)\frac{1}{2}
\ln|z_j-u|\,,
\label{coef}
\end{eqnarray}
where $U=(u|0)$, $V=(v|0)$, $ \hat\kappa_j $ is the same as
in (\ref{funct}), and $\hat X(t,\bar t;t',\bar t')$
is the correlator (\ref{corr}) on the genus-$(n-1)$
supermanifold.
Further,
$ \tilde G_0 $ and every term in $ \tilde B_1 $
represents the genus-0 correlator
for the given points. To calculate
the pre-exponent in (\ref{gint}),
one leaves terms $\geq1$, the estimation
$|z_1-u|\sim|z_2-u|\sim\sqrt{|k|}$ and
$\vartheta_1\sim\vartheta_2\sim|k|^{1/4}$ being used.
In the considered case the pre-exponent does not depend on
$\tilde p_1$ and $\tilde p_2$.  To check (\ref{coef}),
one substitutes (\ref{coef}) in (\ref{gint}) and
calculates the integral, the result being compared
with (\ref{ampl}). In doing so eqs.(\ref{mast}) are used where
the index "1" concerns to  the degenerated handle.

As an example, we consider one-loop amplitude ($n=1$).
Then the integration measure in (\ref{gint}) is
equal to 1, and $\hat X(t,\bar t;t',\bar t')$
is $(\ln|z_1-z_2\vartheta_1\vartheta_2|)/2$.
Product of the two first factors in the pre-exponent
is $O_{(b)}(\{\zeta\})\overline{O_{(b)}(\{\zeta'\})}$ where
the  holomorphic function $O_{(b)}(\{\zeta\})$
can be written down as the sum over 10-vectors
$\tilde\zeta_1$ and $\tilde\zeta_2$ to be polarizations of
bosons carrying momentum $\tilde p_1$ and, respectively,
$\tilde p_2$:
\begin{eqnarray}
O_{(b)}=\sum_{\tilde\zeta_1,\tilde\zeta_2}\biggl[
-\tilde\zeta_1\tilde\zeta_2+\sum_{j=1}^2\tilde\zeta_1
\frac{(\kappa_j+ip_j)\vartheta_j}{2(z_j-u)}
\sum_{l=1}^2\tilde\zeta_2
\frac{1}{2}(\kappa_j+ip_j)\vartheta_j\biggl]
\nonumber\\
\times
\biggl[-\frac{\tilde\zeta_1\tilde\zeta_2}{u-v}+\sum_{r=3}^4
\tilde\zeta_1
\frac{(\kappa_r+ip_r)\vartheta_r}{2(z_r-v)}
\sum_{s=3}^4\tilde\zeta_2
\frac{(\kappa_s+ip_s)\vartheta_s}{2(z_r-u)}
\biggl]-\frac{2}{u-v}\,.
\label{bun}
\end{eqnarray}
Here $\kappa$ is the same as in
(\ref{funct}).
In (\ref{bun}) summation is performed over all
independent polarizations. The term $-2/(u-v)$
appears because from (\ref{z1h}) factor
$8/(u-v)$ appears, but not $10/(u-v)$. This term being the
Faddeev-Popov ghost contribution to the unitarity equations,
just cancels the
non-physical polarizations to (\ref{unit})
(the
proof of this statement will be given in other place).

For the $n$-loop amplitude, factor
$\hat Z_{\tilde L,\tilde L'}^{(n-1)}(\{q,\bar q\} $
\ref{gint}) is equal the genus-$(n-1)$ measure
on the supermanifold formed by all handles, except for
the degenerated handle. As for $n=1 $,
the right part of (\ref{gint}) is reduced to (\ref{unit})
once the integration over $y$, $t_1$, $t_2$ and $t_3=t$
to be performed. One can verify
all the unitarity equations for boson loops,
at least, for massless states (such check will be given
in another place).

\section {Fermion loop}
We derive now the two-particle
unitarity equations for a loop with fermion states
in the holomorphic sector ($l_1=1/2 $).  Then at $k\to0 $
the holomorphic Green  function preserves the singularity at
$z=u$ and $z=v$.  Thus, the function and, therefore,
other functions in (\ref{corr}), differ from
corresponding genus-$(n-1)$ functions.  In particular,
Green function $R_{(0)}(t_1,t_2)$ being
the limit of the appropriate genus-1 one
(for $l_2=0$ and for $2l_2=1$), is given by
\begin{equation}
R_{(0)}(t,t')=\ln(z-z')-\frac{\vartheta\vartheta'}
{2(z-z')}\Biggl[\sqrt{\frac{(z-u)(z'-v)}{z-v)(z'-u)}}+
\sqrt{\frac{(z-v)(z'-u)}{(z-u)(z'-v)}}\,\Biggl]
\label{limgrn}
\end{equation}
For
deriving (\ref{limgrn}) one may, for example, use
genus-1 functions in \cite{danphr}.  The same kind
singularity appears in the higher genus functions.
When all Grassmann moduli are equal to zero it can
be seen from expression \cite{danphr} for the fermion part of
the Green function. Of course, the singularity takes place for
non-zero Grassmann modules, too (see the Appendix).  If $z\to
u$, but $|z'-u|\sim1$, then Green function
$R_L^{(n)}(t,t';\{q\})\equiv R(t,t')$ is given by
\begin{equation}
R(t,t')=R_{(r)}(U,t')-
\frac{\vartheta\phi(t')}{2\sqrt{(z-u)}}\,,
\label{grn}
\end{equation}
where  $R_{(r)}(U,t')$ is its
regular at $z\to u$ part, $U=(u|0)$, and $\phi(t')$ is a
coefficient at the singularity. In this case $\phi(t')$ has no
a singularity at $z'\to u$ due to the  symmetry of the Green
function under the interchange between its arguments.
The kindred formula arises, if $z\to v$.  The
discussed singularities are due to the fermion part of the Green
function, but for non-zero Grassmann moduli, the
singularities arise also for its boson part and, hence, for
scalar functions $J_s^{(n)}(t;\{q\};L)$. If $z\to z'\to u$, then
\begin{equation}
R(t,t')=R_{(0)}(t,t')+R_{(rr)}(U,U)-
\frac{\vartheta\phi(U)}{2\sqrt{(z-u)}}
-\frac{\vartheta'\phi(U)}{2\sqrt{(z'-u)}}
\label{grnuu}
\end{equation}
where $R_{(rr)}(U,U)$ is the regular at $z '\to u $ part.
The $\phi(U)$ is proportional  to
Grassmann moduli.  When any of arguments of $R(t,t')$ go to
$U=(u|\mu=0)$ or to $V=(v|\nu=0)$ (or both they does it) the
Green function has no the discussed singularities since $
\vartheta=0 $.  Then eqs. (\ref{rels}) are trues in the
considered case, too (see the Appendix).
From (\ref{limgrn}) at $z\to z'\to u$
and (\ref{corr}), $\tilde G_0$ in (\ref{gint}) is equal to
\begin{equation}
\tilde G_0=-(\hat\kappa_1+ip_1)
(\hat\kappa_2+ip_2)\biggl
[\frac{1}{2}\ln|z_1-z_2|-
\frac{\vartheta_1\vartheta_2}
{8(z_1-z_2)}\Biggl(\sqrt{\frac{z_1-u}{z_2-u}}+
\sqrt{\frac{z_2-u}{z_1-u}}\Biggl)\Biggl]\,.
\label{tildg}
\end{equation}
If the loop is the fermion one for both movers
($l_1=l_1'=1/2 $), then the  expression
in square brackets is added by the term, which is a
complex conjugated to last term in the above brackets.
Other coefficients for the exponents in (\ref{gint}) are given
by (\ref{coef}) where $\hat X(t,\bar t;t',\bar t')$
it is the limit at $k\to0 $ of the vacuum
correlator (\ref{corr}).  Hence at $z\to u$,
the correlator $\hat X(t,\bar t;t',\bar t')$ in
(\ref{coef})  is given by
\begin{equation}
\hat X(t,\bar t;t',\bar t')=\hat
X_{(r)}(U,\bar U;t',\bar t')-
\frac{\vartheta\varphi(t')}{2\sqrt{(z-u)}}\,.
\label{corlim}
\end{equation}
At
$z\to u$ and $z'\to u$, it is equal  to
\begin{equation}
\hat X(t,\bar t;t',\bar t')=\hat
X_{rr}(U,\bar U;
U,\bar  U)-
\frac{\vartheta\varphi(U)}{2\sqrt{(z-u)}}
-\frac{\vartheta'\varphi(U)}{2\sqrt{(z'-u)}}
\label{corlm}
\end{equation}
where $\varphi(t)$ has no a singularity at
$z=u$.  Further, $\phi(U)$ is proportional
Grassmann moduli.  In so doing $\varphi(t)$ depends not only
on $t$, but also on $\bar t$. If $l_1'=1/2$ then discussed
singular terms need to be added by their complex conjugated.
The discussed expressions for factors in (\ref{gint}) are
checked by the calculation of integral (\ref{gint}) after
the substitution of (\ref{coef}), (\ref{tildg}) and of the
expressions given below for the pre-exponent.

In particular, $\hat Z_{\tilde L,\tilde L'}^{(n-1)}
(\{q,\bar q\}$ in (\ref{gint}) is given by
(\ref{hol}) with $Z_L^{(n)}(\{q\})$
to be replaced by\footnote{
The
multiplier 1/16 in the next expression in the text
presents since the
multiplier 16 in (\ref{z1h}) at $2l_1=1$ arises from the trace
of the Dirac matrix product in the fermion loop.}
$kZ_L^{(n)}(\{q\})/16$ at $k=0$.
In addition,
$\omega^{(n)}(\{q\},L)$ is replaced by $
\tilde\omega^{(n-1)}(\{q\},\tilde L)$, whose elements are
equal to elements $[\omega^{(n)}(\{q\},L)]_{jl}$ for
$j,l=2,\dots,n$ at $k=0 $. As in
(\ref{mast}), indexes $j,l=2,\dots,n$ are assigned to
non-degenerated handles. If $l_1'=1/2 $, then $Z_{L '}^{(n)}
(\{q\})$ is replaced by $kZ_{L'}^{(n)}
(\{q\})/16$ at $k=0$, and $\omega^{(n)}
(\{q\},L')$  is replaced by $\tilde\omega^{(n-1)}
(\{q\},\tilde L)$, its elements being $[\omega^{(n)}
(\{q\},L')]_{jl}$ at $k=0$. If $l_1'=0$, then
$Z_{L'}^{(n)}(\{q\})$ and $\omega^{(n)}(\{q\},L')$
both are replaced by the corresponding functions on the
genus-$(n-1)$ supermanifold.
At last,
$O_{\tilde L}(\{p,\zeta\},\tilde p_1,\{t\},\{q\})$
is given by
\begin{equation}
O_{\tilde L}(\{p,\zeta\},\tilde p_1,\{t\},
\{q\})=\exp\biggl[\sum_{j=1}^2
\frac{(2\eta_j\zeta_{(j)}+ip_j\vartheta_j)}
{8\sqrt{z_j-u}}\hat\Psi
\biggl]\,,
\label{ofact}
\end{equation}
where $\eta_j$ is as in (\ref{funct}), and the
10-vector $\hat\Psi$ depending on the final state parameters,
is
\begin{equation}
\hat\Psi=\sum_{j=3}^4[2\eta_j\zeta_{(j)}D(t_j)+ip_j\vartheta_j]
\varphi(t_j)+i\tilde p_1\varphi(V)+i\tilde p_2\varphi(U)\,,
\label{hpsi}
\end{equation}
where
$D(t)$ is given by (\ref{supdr}) and $\varphi(t)$ is defined
in (\ref{corlim}).  Like $\varphi(U)$, the function
$\varphi(V)$ is proportional the Grassmann moduli. If $l_1'=1/2$,
then expression for conjugated factor in (\ref{gint}) is
calculated in a similar kind. If $l_1'=0$, then it is the same
as for the boson loop. As it was already noted, all the
expressions above are checked by their substitution in
(\ref{gint}) with the following calculation of the integral over
$\tilde p_1$, the result being compared with (\ref{ampl}). The
desired $n$-loop amplitude $ \tilde A^{(n)}$ with two Ramond
states is
\begin{equation}
\tilde A^{(n)}=\bar\psi(\tilde
p_2)(\tilde T_3^{(n)}+ \tilde T_1^{(n)})\psi(\tilde p_1)\,,
\label{ramp}
\end{equation}
where
$\psi(p)$ is the Majorano-Weyl spinor, which satisfies to
$\Gamma_{11}\psi(p)=\psi(p)$ and to the Dirac equation
$(\Gamma p)\psi(p)=0 $. Here $\Gamma^M$ is the Dirac matrix,
and $\Gamma_{11}$ is the product of all
of ten matrixes above. Furthermore, $\tilde T_1^{(n)}$
contains one Dirac matrix, and $\tilde T_3^{(n)}$
contains the anti-symmetrized product of three matrixes
discussed.  Hence
\begin{equation}
O_{\tilde L}(\{p,\zeta\},\tilde p_1,\{t\},
\{q\})=trace[(T_3^{(0)}+
T_1^{(0)})(\Gamma\tilde p_1)(T_3^{(n-1)}+
T_1^{(n-1)})(\Gamma\tilde p_2)]+\dots
\label{rfc}
\end{equation}
where dots denote terms, which  are nullified once the
integrations in (\ref{ampl}) and
summation over the spin structures
are performed. Further, $T_1^{(n)}$ contains one Dirac matrix,
and $T_3^{(n)}$ contains the anti-symmetrized product of three
such matrixes.  As far as 10-spinor in (\ref{ramp}) satisfy to
the Weyl condition, the trace of the unity
matrix is equal to 16.

Eq.(\ref{ofact}) is a 4-degree polynomial in the exponent
since the exponent is the sum of products of Grassmann
variables. Odd degrees  disappear once Grassmann integration
is performed. Even
degrees being represented as (\ref{rfc}), determine
the integrand for
$ \tilde A^{(0)}$ and $\tilde A^{(n-1)}$. For this purpose
(\ref{hpsi}) is re-written as
\begin{equation}
\hat\Psi=\Psi-iP\Phi +i\tilde p
[\varphi(V)-\varphi(U)]\,,
\label{fact}
\end{equation}
where $P=p_1+p_2 =\tilde p_1+\tilde p_2$, $\tilde p=\tilde
(p_1-\tilde p_2)/2$,
\begin{equation}
\Phi=\frac{1}{2}[\varphi(t_3)+\varphi(t_4)-
\varphi(U)-\varphi(V)]\,,
\label{fct}
\end{equation}
\begin{equation}
\Psi=2\eta_3\zeta_{(3)}D(t_3)\varphi(t_3)
+2\eta_4\zeta_{(4)}D(t_4)\varphi(t_4)
+i(p_3-p_4)\frac{1}{2}[\varphi(t_3)-\varphi(t_4)]
\,.
\label{psi}
\end{equation}
Function $\varphi(t)$ is defined in (\ref{corlim}).
In that case (\ref{ofact}) appears to be
\begin{eqnarray}
O_{\tilde L}(\{p,\zeta\},\tilde p_1,\{t\},
\{q\})=\exp\Biggl[\sum_{r=1}^2
\frac{4\eta_r\zeta_{(r)}+i(p_1-p_2)
(\vartheta_1-\vartheta_2)}{16\sqrt{z_r-u}}\Psi
\nonumber\\
-\sum_{j=1}^2
\frac{2i\eta_j(P\zeta_{(j)})+(p_1p_2)
(\vartheta_1+\vartheta_2)}{8\sqrt{z_j-u}}\Phi
+i\sum_{j=1}^2\frac{\vartheta_1+\vartheta_2}{16\sqrt{z_j-u}}
(P\tilde\Psi)
\nonumber\\
+ \sum_{j=1}^2
\frac{4\eta_s(\zeta_{(s)}\tilde p)+i(p_1-p_2)\tilde p
(\vartheta_1-\vartheta_2)}{16\sqrt{z_s-u}}
[\varphi(V)-\varphi(U)]\Biggl]\,.
\label{fun}
\end{eqnarray}
The tree amplitude is calculated from the one -loop case where
Grassmann moduli is absent.
Representing the factor in (\ref{gint}) as
(\ref{rfc}), one obtains that
\begin{eqnarray}
4\sqrt2T_1^{(0)}=-
\biggl(\Gamma[(\zeta_3\zeta_4)\frac{1}{2}(p_3-p_4)-
(\zeta_4p_3)
\zeta_3+(\zeta_3p_4)\zeta_4]\biggl)
\frac{\eta_3\vartheta_3\eta_4\vartheta_4}{2(z_3-z_4)}
\nonumber\\
\times
\biggl[\frac{v-u}{(z_3-u)(z_4-v)}+
\frac{v-u}{(z_3-v)(z_4-u)}\biggl]
+\frac{[(\zeta_3(\tilde p_3-\tilde p_4)\zeta_4+
(\zeta_4(\tilde p_3-\tilde p_4)\zeta_3](v-u)^2}
{4(z_3-u)(z_3-v)(z_4-u)(z_4-v)}\,,
\nonumber\\
4\sqrt2T_3^{(0)}=\frac{[(\Gamma\zeta_3)(\Gamma\zeta_4)
(\Gamma(p_3-p_4))]_a\eta_3\vartheta_3\eta_4\vartheta_4(v-u)^2}
{4(z_3-u)(z_3-v)(z_4-u)(z_4-v)}\,,
\label{tzer}
\end{eqnarray}
where $[...]_a$ denotes the anti-symmetrized
(with $1/6$) expression. In (\ref{tzer}) terms omitted, which are
nullified once the Grassmann integrations are performed.  As
above, $v$ or $u$ is the coordinate of the fermion emission
vertex with momentum $\tilde p_1$ and, respectively, $\tilde
p_2$. For the amplitude with  boson momenta to be $p_1 $ and
$p_2 $, the appropriate expressions arise as the limit
$u\to\infty $ of (\ref{tzer}) with the following replacement
$v\to u$ along with the replacement of indexes $3\to1$ and
$4\to2$. Being substituted in (\ref{gint}),
eqs.(\ref{tzer}) give (up to the normalization)
the amplitude obtained in
\cite{fried}. To derive (\ref{tzer}),
one integrates by pert
those terms in
(\ref{gint}), which are
proportional $
\eta_1\eta_2 p_1p_2/[(z_1-u)(z_1-z_2)]$ and
$\eta_1\eta_2p_1p_2/[(z_2-u)(z_1-z_2)]$.
The first term is
integrated over $z_2$, and the second one is integrated over
$z_1$.  Due to (\ref{tildg}) for
$\tilde G_0$, the first term turns to $-\eta_1\eta_2
p_2\tilde p_1/[(z_1-u)(z_2-u)]$, and the second one appears to
be $\eta_1\eta_2 p_1\tilde p_1/[(z_1-u)(z_2-u)]$.
Being multiplied by the $1/(z_3-z_4) $ factor,
expression in square brackets in first of eqs. (\ref{tzer})
is represented either as
\begin{equation}
\frac{2(v-u)}{(z_4-u)(z_4-v)(z_3-z_4)}\,
-\,\frac{(v-u)^2}{(z_3-u)(z_4-u)(z_4-v)}\,
-\,\frac{(v-u)^2}{(z_3-v)(z_4-v)(z_4-u)}\,,
\label{mst1}
\end{equation}
or that obtained by replacements
$z_3\rightleftharpoons z_4$ and $v\rightleftharpoons u$.
When
(\ref{mst1}) is multiplied by $\eta_3\eta_4p_3p_4$,
the first term in (\ref{mst1}) is integrated by parts
over $z_3$. Then it appears to be
proportional to
\begin{equation}
-\frac{2\eta_3\eta_4
(v-u)}{(z_4-u)(z_4-v)} \Biggl[\frac{p_3\tilde p_1}{z_3-v}+
\frac{p_3\tilde p_2}{z_3-u}\Biggl]\,.
\label{mst2}
\end{equation}
It is useful to note that from (\ref{limgrn}),
in the considered $n=1$ case the function $ \varphi (t) $ in
(\ref{fun}) is given by
\begin{equation}
\varphi(t)=-\vartheta\sqrt{\frac{u-v}{(z-u)(z-v)}}\,.
\label{ph0}
\end{equation}
For the arbitrary $n$ (including $n=0 $)
the term
$T_3^{(n-1)}$ in (\ref{rfc}) turns out from
the term $\sim\Psi_{M_1}\Psi_{M_2}\Psi_{M_3}\Phi $, which
arises in the expansion over the exponent in (\ref{ofact}) from
product of the third degree of the sum on $s$ in (\ref{ofact})
by $\sim\Phi$. Here $\Psi_M $ is 10-component of
$\Psi$.  Being proportional to $(\tilde p_1\tilde p_2)$, the
discussed terms gives
$ \sim (\tilde
p_1\tilde p_2) $ terms in (\ref{rfc}), which are due to
$trace [T_3^{(0)}(\Gamma\tilde p_1)T_3^{(n-1)}
(\Gamma\tilde p_1)]$. As far as spinor structure in
$T_3^{(n-1)}$ is independent from the intermediate state
variables, $T_3^{(n-1)}$  is calculated in the unique way as
\begin{equation}
T_3^{(n-1)}=\frac{i}{48}(\Gamma\Psi)(\Gamma\Psi)(\Gamma\Psi)
\Phi\,.
\label{t3t}
\end{equation}
As far as $T_3^{(n-1)}$ and $T_3^{(0)}$ both are
anti-symmetrized in Dirac matrices, in the discussed {\it
trace}, there is no terms proportional to scalar products of
those vectors, which both belong to $T_3^{(0)}$ or
to $T_3^{(n-1)} $.  For the same reason, there are no terms
where each of $\tilde p_1$ and $\tilde p_2$ in (\ref{rfc})
forms the scalar product with the vector belonging to the same
$T_3^{(0)}$ or $T_3^{(n-1)}$. Other terms in the {\it trace} may
arise only from those terms in the expansion of the exponent
(\ref{ofact}), which contain a square of the sum over $r$. If
$T_3^{(n-1)}$ is given by (\ref{t3t}), the quadratic in
$(\tilde p_1+\tilde p_2)$ part of the {\it trace} is obtained,
too. The
remaining terms are  quadratic in
$(\tilde p_1-\tilde p_2)$.  They
do not contained in the expansion of the exponent
(\ref{ofact}) since they do not include
$\sim\Psi\Psi\Psi\Phi$.  Hence, for
the corresponding terms to be in the unitarity equation for
the boson amplitude, the  following integration
equation must be satisfied
\begin{equation}
<\Psi_M\Psi_N\Biggl[1-\frac{1}{16}
[\varphi(V)-\varphi(U)]\biggl(P^2\Phi+i(P\Psi)\biggl)\Biggl]-
\frac{i}{4}\Psi_M\Psi_N(\Psi\tilde p)\Phi>=0\,,
\label{constr1}
\end{equation}
where $ <...>$ means the integration of the explicit expression
in (\ref{constr1}) (for details, see eq.(\ref{ampr}) and
the text below), which is previously multiplied by the following
from (\ref{gint}) factors, the summation over $\tilde L,\tilde
L')$ being performed.  The integral is calculated over
variables assigned to $\tilde
A^{(n-1)}$.  To calculate
$T_1^{(n-1)}$ in (\ref{rfc}), we consider $trace
[T_3^{(0)}(\Gamma\tilde p_1)T_1^{(n-1)} (\Gamma\tilde p_1)]$.
Extracting it from terms in (\ref{ofact}), we obtain that
\begin{equation}
T_1^{(n-1)}=-i(\Gamma\Psi)\Phi+
\frac{i}{8}(\Gamma\Psi)(P\Psi)[\varphi(V)-\varphi(U)]
\Phi\,,
\label{t1t}
\end{equation}
where notations are the same as in (\ref{ofact}).
With (\ref{t1t}) and (\ref{constr1})
to be taken into account,
the remaining linear
$\Psi$ terms in (\ref{ofact}) originate
all remaining terms in (\ref{rfc}), except that, which is
proportional to $trace[(\Gamma\tilde p)T_1^{(0)}]$
times $trace[(\Gamma\tilde p)T_1^{(n-1)}]$. In addition, it
can not be originated by the unity in
(\ref{ofact}) since it does not contain $\Psi$.
Hence the integration relation arises to be
\begin{equation}
<16+trace\biggl[\frac{1}{2}(\Gamma T_1^{(n-1)})
(\Gamma\tilde p)\biggl]+\sqrt2 [\varphi(V)-\varphi(U)][P^2\Phi+
(\Psi P)>=0\,,
\label{constr2}
\end{equation}
where, as in (\ref{constr1}), $<...>$ means
integration of the explicit expression in (\ref{constr1})
multiplied by  all following from
(\ref{gint}) factors, the summation over $(\tilde L,\tilde L')$
being performed.
Other notations are the same as in (\ref{ofact}).

Thus, the $(n-1)$-loop amplitude
$\tilde A^{(n-1)}$ for the transition of two massless
Ramond states (carrying  $\tilde p_1$ and $\tilde
p_2$) to two massless bosons with momenta and polarizations to be
$(p_3,\zeta_{(3)})$ and $(p_4,\zeta_{(4)})$,  is
given by
\begin{eqnarray}
A^{(n-1)}=\frac{2g^{2n}}{2^n(n-1)!}\int
|(z_4-u)(z_4-v)|^2\sum_{\tilde L,\tilde L'}
\hat Z_{\tilde L,\tilde L'}^{(n-1)}(\{q,\bar q\})
\tilde F_{\tilde L,\tilde L'}^{(n-1)}
\nonumber\\
\times
{\cal O}_{\tilde L}\overline{{\cal O}_{\tilde L'}}
(dqd\bar qdtd\bar t)' \,,
\label{ampr}
\end{eqnarray}
where $g $ is a coupling constant,
and summation is performed over
super-spin structures on the genus-$(n-1)$ supermanifold.
The sum includes odd superspin structures due to
the contribution to the unitarity from the fermion loop
with $l_2=1/2 $.
The integrand in (\ref{ampr}) is expressed
through functions determining the $n$-loop amplitude
(\ref{ampl}) in the region where $k\to0$. In doing so $\hat
Z_{\tilde L,\tilde L'}^{(n-1)}(\{q,\bar q\})$  is
calculated through $Z_{L, L'}^{(n)}(\{q,\bar q\})$ as
it was explained next eq.(\ref{corlm}).
Among other things, it
depends on coordinates $v$ and $u$ of the fermion vertices.
Further, $\tilde F_{\tilde L,\tilde L'}^{(n-1)}$ is
calculated through $\hat F_{\tilde L,\tilde L'}^{(n-1)}(\eta_1,
\eta_2, \bar\eta_1, \bar\eta_2)$, which is the expression in
(\ref{funct}) integrated over $(\eta_3,\bar\eta_3)$
and $(\eta_4,\bar\eta_4)$. In this case the genus-$n$ correlator
(\ref{corr}) is replaced by
$\hat
X(t,\bar t;t',\bar t')$ to be the above correlator at
$k\to0$. Besides, $
(z_1|\vartheta_1)\to (v|0)$, $(z_2|\vartheta_2)\to(u|0)$,
$p_1\to\tilde p_1$ and $p_2\to\tilde p_2 $.  If
$\tilde p_1$ and $\tilde p_2 $ are carried by the Ramond
bosons, then
\begin{equation}
{\cal O}_{\tilde L}=
\bar\psi(\tilde p_2)(T_3^{(n-1)}+
T_1^{(n-1)})\psi(\tilde p_1)
\label{ferm}
\end{equation}
where $\psi(p)$ is the spinor, and other
definitions are given in (\ref{t3t}) and (\ref{t1t}).
In this case $\overline{{\cal O}_{\tilde L'}}$ is calculated in
the like manner. In doing so $\tilde F_{\tilde L,\tilde
L'}^{(n-1)}$ in (\ref{ampr}) is equal to $\hat F_{\tilde
L,\tilde L'}^{(n-1)} (\eta_1,\eta_2,\bar\eta_1,\bar\eta_2)$ at
$\eta_1=\eta_2=0$ and $\bar\eta_1=\bar\eta_2=0$. So
\begin{equation}
\tilde F_{\tilde
L,\tilde L'}^{(n-1)}= \hat F_{\tilde
L,\tilde L'}^{(n-1)}(0,0,0,0)\,.
\label{tilf})
\end{equation}
If
$\tilde p_1$ and $\tilde p_2$ are carried by the Ramond
fermions, then either holomorphic pare, or
the anti-holomorphic one of the state
is described by boson wave function. If it is the holomorphic
part, then
${\cal O}_{\tilde L}=1$, and
\begin{equation}
\tilde F_{\tilde
L,\tilde L'}^{(n-1)}= \int d\eta_1d\eta_2\hat F_{\tilde
L,\tilde L'}^{(n-1)}(\eta_1,\eta_2,0,0)\,.
\label{tilb}
\end{equation}
Further, $\overline{{\cal O}_{\tilde L'}}$ is calculated with
(\ref{ferm}). The
fermion vertex coordinates are fixed to be $(v|0)$ and $(u|0)$,
along with the boson local coordinate $z_4$. Since (\ref{ampr})
has $SL(2)$-symmetry, one could fix any other $(3|2)$ variables,
but we do not develop this topic here.

It can be checked that with
(\ref{constr1}) and (\ref{constr2}), the amplitude
(\ref{ampr}) for longitudinal boson emission vanishes as it is
required by gauge invariance (to be shown otherwise).

Expressions (\ref{constr1}) and (\ref{constr2})
are given by (\ref{ampr}) with replacing $ {\cal O}_{\tilde L}$
(or $\overline{{\cal O}_{\tilde L'}}$, or
both them) by functions inside $ <...> $ (or their complex
conjugated). At present we check the discussed relations
only for the tree approximated amplitude (\ref{ampr}).
Then the first multiplier under the sum symbol
in (\ref{ampr}) is equal to unity, and other factors are
calculated by
(\ref{tildg}) and (\ref{ph0}). For the checking, it is
convenient to fix $u\to\infty$ or $v\to\infty$.
If
$u\to\infty$, then terms with $\eta_3\eta_4
p_3p_4/[(z_4-v)(z_3-z_4)]$ are integrated by parts, like the
expressions discussed after eq.(\ref{tzer}). Of course,
the check of the relation can be performed for any $u$ and
$v$.  In that case eq.(\ref{mst1}) is used with the
integration by parts leading to
(\ref{mst2}).  The verifying the discussed relations for loop
amplitudes needs an additional study.

Work is maintained by grant No. RP1-2108 of the U.S. Civilian
Research and Development Foundation for the Independent States
of the Former Soviet Union (CRDF), and also grants No.
00-02-16691 and No. 00-15-96610 of the Russian Funds for basic
Researches.

\def\thesection{Appendix \Alph{section}}
\def\theequation{\Alph{section}.\arabic{equation}}
\setcounter{equation}{0}

\appendix
\section {Green functions}
Here we argue relations
(\ref{rels}) for the Ramond sector and clarify formulas
of Section 5 for the vacuum correlator.
For this purpose
we use eq.\cite{prepr} of the
genus-$n$ Green function through functions of genera-$n_i$ where
$\sum_i n_i=n$. In this case $n$ handles are divided into groups
where the $i$-th group consists from the $n_i$ handles. Each
group is given by the set of $\{q\}_i$ parameters of the
super-Schottky group.  For simplicity we consider all the
super-spins structures $L_i$ to be even ($L =\{L_i\}$).
Generally, to construct the Green function, one needs solely to
build an expression, which satisfies known relations under
transformations $t\to t_s^b$ and $t\to t_s^a$ for $2\pi$-twist
about $B_s$ and $A_s$-cycle:
\begin{eqnarray}
R_L^{(n)}(t_s^b,t';\{q\})=R_L^{(n)}(t,t';\{q\})+
J_s^{(n)}(t';\{q\};L)\,,
\nonumber\\
R_L^{(n)}(t_s^a,t';\{q\})=R_L^{(n)(r)}(t,t';\{q\})\,,
\label{rtrans}
\end{eqnarray}
where $R_L^{(n)(s)}(t,t';\{q\})$ is obtained by $2\pi$-twist
about the corresponding Schottky circle (if the circle is
assigned to the Ramond type handle, the Green function has the
square cut).  Simultaneously, in doing so one determines the
scalar functions. Furthermore, one determines the period matrix
using the relations for these functions
under
transformations $t\to t_s^b$ and $t\to t_s^a$ for $2\pi$-twist
about $B_s$ and $A_s$-cycle. The above relations are
\begin{eqnarray}
J_r^{(n)}(t_s^b;\{q\};L) = J_r^{(n)}(t;\{q\};L)+2\pi i
\omega_{sr}(\{q\},L)\,,
\nonumber\\
J_r^{(n)}(t_s^a;\{q\};L)=J_r^{(n)(s)}(t;\{q\};L)+
2\pi i\delta_{rs}\,.
\label{trjs}
\end{eqnarray}
In
doing so we consider $K_L^{(n)}(t,t';\{q\})$ defined to be
\begin{equation}
K_L^{(n)}(t,t';\{q\})=D(t')R_L^{(n)}(t,t';\{q\})
\label{kr}
\end{equation}
where the spinor derivative is defined by (\ref{supdr}).
We build (see \cite{danphr,prepr}) a
matrix operator $\hat K=\{\hat K_{sr}\}$ where $\hat K_{sr}$ is
an integral operator vanishing at $s=r$. For $s\neq r$, the
kernel of $\hat K_{sr}$ is $\tilde
K_{L_s}^{(n_s)}(t,t';\{q\}_s)dt'$. Here $\tilde
K_{L_s}^{(n_s)}(t,t';\{q\}_s)$ is related by (\ref{kr}) with
$\tilde R_{L_s}^{(n_s)}(t,t';\{q\}_s)$, which is the
non-singular part (\ref{lim}) of the Green function. So
\begin{equation}
K_{L_s}^{(n_s)}(t,t';\{q\}_s)=
\frac{\vartheta-\vartheta'}
{z-z'}+\tilde K_{L_s}^{(n_s)}(t,t';\{q\}_s)\,.
\label{polk}
\end{equation}
We define kernels together with the
differential $dt'=dz'd\vartheta'/2\pi i$. The discussed operator
performs the integration with $\tilde
K_{L_s}^{(n_s)}(t,t';\{q\}_s)$ over $t'$ along $C_r$-contour,
which surrounds the limiting points associated with the
considered group $r$ of the handles and the cuts between
limiting points for the Ramond handles.  The desired relation
for the Green function is derived in Section 5 of \cite{prepr}.
For every case of interest, the expression obtained can be
transformed to the relation, which allows to verify
(\ref{rtrans}) under the transformations assigned to the $r$-th
group of the handles (Section 5 of \cite{prepr}).
To give the desired expression, we define
$[(1-\hat K)^{-1}\hat K]_{rs}(t,t_1)dt_1$ to be the kernel of
the operator
\begin{equation}
(1-\hat K)^{-1}\hat K=
\hat K+\hat K\hat K+\dots
\label{opr}
\end{equation}
The desired Green function is given by
\begin{eqnarray}
R_L^{(n)}(t,t';\{q\})=
R_{L_r}^{(n_r)}(t,t';\{q\}_r)+
\sum_{s\neq r}\int_{C_s}
K_{L_r}^{(n_r)}(t,t_1;\{q\}_r)
\tilde R_{L_s}^{(n_s)}(t_1,t';\{q\}_s)dt_1
\nonumber\\
+\sum_{p\neq r}\sum_{s}
\int_{C_p}K_{L_r}^{(n_r)}(t,t_1;\{q\}_r)dt_1\int_{C_s}
[(1-\hat K)^{-1}\hat
K]_{ps}(t_1,t_2)dt_2 \tilde R_{L_s}^{(n_s)}(t,t';\{q\}_s)
\label{rpart}
\end{eqnarray}
where $R_{L_r}^{(n_r)}(t,t';\{q\}_r)$ and
$K_{L_r}^{(n_r)}(t,t_1;\{q\}_r)$ are total Green functions
including the singular term in (\ref{lim}) and (\ref{polk}).
The scalar function
$J_{j_r}^{(n)}(t;\{q\};L)$ associated with the $j_r$ handle of
the $r$-th supermanifold appears to be \cite{prepr}
\begin{eqnarray}
J_{j_r}^{(n)}(t;\{q\};L)=J_{j_r}^{(n_r)}(t;\{q\}_r;L_r)+
\sum_{s\neq r}\int_{C_s}D(t_1)
J_{j_r}^{(n_r)}(t_1;\{q\}_r;L_r)dt_1
\tilde R_{L_s}^{(n_s)}(t_1,t;\{q\}_s)
\nonumber\\
+\sum_{p\neq r}\sum_{s}
\int_{C_p}D(t_1)
J_{j_r}^{(n_r)}(t_1;\{q\}_r;L_r)dt_1\int_{C_s}
[(1-\hat K)^{-1}\hat
K]_{ps}(t_1,t_2)dt_2 \tilde R_{L_s}^{(n_s)}(t_2,t;\{q\}_s)\,.
\label{tjr}
\end{eqnarray}
The period
matrix is calculated from (\ref{tjr}) presented to be
\begin{eqnarray}
J_{j_r}^{(n)}(t;\{q\};L)-J_{j_r}^{(n)}(t_0;\{q\};L)=
J_{j_r}^{(n_r)}(t;t_0\{q\}_r;L_r)
+\int_{C_s'}D(t_1)J_{j_r}^{(n_r)}
(t_1;\{q\}_r;L_r)
dt_1
\nonumber\\
\times
R_{L_s}^{(n_s)}(t_1,t;t_0;\{q\}_s)
\nonumber\\
+\sum_{p\neq r}
\int_{C_p}D(t_1)
J_{j_r}^{(n_r)}(t_1;\{q\}_r;L_r)dt_1\int_{C_s'}
[(1-\hat K)^{-1}\hat
K]_{ps}(t_1,t_2)dt_2R_{L_s}^{(n_s)}(t_2,t;t_0;\{q\}_s)
\label{jar}
\end{eqnarray}
where $t_0$ is a fixed parameter. Both
$z$ and $z_0$ lay inside the $C_s'$ contour, and
\begin{eqnarray}
R_{L_s}^{(n_s)}(t_1,t;t_0;\{q\}_s)=
R_{L_s}^{(n_s)}(t_1,t;\{q\}_s)-
R_{L_s}^{(n_s)}(t_1,t_0;\{q\}_s)\,,
\nonumber\\
J_{j_r}^{(n_r)}(t;t_0\{q\}_r;L_r)=
J_{j_r}^{(n_r)}(t;\{q\}_r;L_r)
-J_{j_r}^{(n_r)}(t_0;\{q\}_r;L_r)
\label{jar1}
\end{eqnarray}
where $R_{L_s}^{(n_s)}(t_1,t;\{q\}_s)$ is the total Green
function (\ref{lim}) including the singular term. To prove
(\ref{jar}), one, using (\ref{opr}), calculates the
contribution from the $\ln[(z_2-z-\vartheta_2\vartheta)/
(z_2-z_0-\vartheta_2\vartheta_0)]$ term due to the singularity
of the Green function.  The corresponding integral is
transformed to the one along the cut between
$z_2=z-\vartheta_2\vartheta$ and
$z_2=z_0-\vartheta_2\vartheta_0$. Then it is found to be
\begin{equation}
\int
D(t_2)f(t_2)[\theta(z_2-z-\vartheta_2\vartheta)-
(z_2-z_0-\vartheta_2\vartheta_0)]dz_2d\vartheta_2=
f(t)-f(t_0)
\label{calom}
\end{equation}
where $f(t)$ denotes either the Green function, or
$J_{j_r}^{(n_r)}(t_1;\{q\}_r;L_r)$. As the results, one obtains
(\ref{tjr}). From (\ref{jar}),
the $\omega_{j_rj_s}^{(n)}(\{q\};L)$ element of the period
matrix is found to be
\begin{eqnarray}
2\pi
i\omega_{j_rj_s}^{(n)}(\{q\};L)=\delta_{j_rj_s}\ln k_{j_r}
+(1-\delta_{j_rj_s})\int_{C_s}D(t)
J_{j_r}^{(n_r)}(t;\{q\}_r;L_r)dt
J_{j_s}^{(n_s)}(t;\{q\}_s;L_s)
\nonumber\\
+\sum_{p}\int_{C_p}D(t)
J_{j_r}^{(n_r)}(t;\{q\}_r;L_r)dt\int_{C_s}[(1-\hat K)^{-1}\hat
K]_{pr}(t,t')dt'
J_{j_s}^
{(n_s)}(t';\{q\}_s;L_s)\,.
\label{omjr}
\end{eqnarray}
Now we use the above expressions in the case of two group of the
handles present. The first group consists of solely the
degenerated handle, and the second group consists of all the
remaining ones. The genus-1 scalar function is given by
\begin{equation}
J(t)=\ln\frac{z-u}{z-v}\,.
\label{sfun}
\end{equation}
To receive the first relation in
(\ref{rels}) for the boson loop, we substitute eq.(\ref{sfun})
in (\ref{omjr}) for $\omega_{11}$
and calculate (\ref{omjr}) at $k\to0 $.  In
this case the integration contour over $t_1 $ rounds
the non-degenerated handles, but it can be transformed to a
contour round the poles of $J(z)$. Since for $k\to0$ the
remaining part of the integrand  has no singularities at
$z=(u,v)$ and $z'=(u,v)$, the integral is  the sum over the
residues in these points. With (\ref{rpart}), the desired
relation appears.  To see the second relation in (\ref{rels}),
one uses (\ref{omjr}) for $\omega_{1l}$ where the integration
contour over $t$ rounds the non-degenerated handles,
(\ref{sfun}) being substituted. At $k\to0$ the integral
(\ref{omjr}) over $t$  is given by the sum over the residues in
$z=u $ and $z=v $. With (\ref{jar1}) the desired relation
appears. They relation for $J_1 $ is obtained by substitution
of (\ref{sfun}) in (\ref{jar1}) for $J_1 $ and the calculation
it at $k\to0$. The integral is equal to the sum over the
residues in $z=u $ and $z=v $. With (\ref{rpart}), the desired
relation appears.  To receive (\ref{grn}), we use (\ref{rpart})
where the integration contour over $t_1$ rounds the degenerated
handle.  The first factor in (\ref{rpart}) is calculated  by
(\ref{kr}) for Green's function (\ref{limgrn}).  The regular at
$z\to u$  part in (\ref{grn}) is given by the contribution from
the pole at $z_1=z=u $ originated by the logarithmic term in
(\ref{limgrn}).  Eq.(\ref{grnuu}) and formulas for scalar
functions are derived in the similar way.  As the result, with
(\ref{corr}), expressions (\ref{corlim}) and (\ref{corlm})
appear.

 {}

\newpage

\thispagestyle{empty}

\vspace*{8cm}
\noindent {\bf G.S.Danilov}

\bigskip
\begin{center}
{\Large {\bf MANIFEST CALCULATION AND THE FINITENESS
OF THE SUPERSTRING FEYNMAN DIAGRAMS
}}
\end{center}

\end{document}